\documentclass[amsmath,amssymb]{revtex4}
\usepackage{graphicx}% Include figure files
\usepackage{bm}% bold math
\newcommand{\pp}[1]{\phantom{#1}}
\newcommand{\be}{\begin{eqnarray}}
\newcommand{\ee}{\end{eqnarray}}

\newcommand{\Tr}{{\,\rm Tr\,}}

\newcommand{\ba}{\begin{array}}
\newcommand{\ea}{\end{array}}

\begin{document}
\title{
Notes on Geometric-Algebra Quantum-Like Algorithms}
\author{Diederik Aerts $^1$ and Marek Czachor $^{1,2}$}
\affiliation{
$^1$ Centrum Leo Apostel (CLEA) and Foundations of the Exact Sciences (FUND)\\
Vrije Universiteit Brussel, 1050 Brussels, Belgium\\
$^2$ Katedra Fizyki Teoretycznej i Informatyki Kwantowej\\
Politechnika Gda\'nska, 80-952 Gda\'nsk, Poland}

\begin{abstract}
In these notes we present preliminary results on quantum-like algorithms where tensor product is replaced by  geometric product. Such algorithms possess the essential properties typical of quantum computation (entanglement, parallelism) but employ additional algebraic structures typical of geometric algebra --- structures absent in standard quantum computation. As a test we reformulate in Geometric Algebra terms the Deutsch-Jozsa problem.
\end{abstract}
\maketitle

\section{Introduction}

Quantum algorithms are based on tensor products. Common wisdom states that tensor products are needed for entanglement. A similar situation was encountered in the early 1990s in connectionist systems, and led to the paradigm known as tensor product representations \cite{Smolensky}. However, nowadays the cognitive science community seems to depert from tensor product representations in favor of their ``compressed forms" such as Binary Spatter Codes (BSC) 
\cite{Kanerva} or Holographic Reduced Representations (HRRs) \cite{Plate}. 

The main reason why BSC and HRRs replace tensor product representations is that tensor multiplication expands dimensions of the associated liner spaces (tensor product of 
two $n$-tuples is an $n^2$-tuple). HRRs, for example, replace tensor product with circular convolution, an operation that does not change the dimension (circular convolution of two $n$-tuples is again an $n$-tuple). Circular convolution is often referred to as a compressed form of the tensor product. Similarly, in BSC one replaces tensor products by XORs of binary strings. Quite recently, following the general program of investigating similarities and differences between AI, semantic analysis, and quantum information \cite{AC},  we have reformulated BSC in terms of Geometric Algebra (GA) \cite{ACDM,GA1,GA2}. This reformulation was made possible by the observation that XOR has a natural representation at the level of geometric product. 

In these notes we present a similar reformulation of the Deutsch-Jozsa algorithm \cite{DJ}. As one can see, after minor modifications the GA algorithm works analogously to the quantum one. Accordingly, it is possible that GA algorithms can perform more general tasks until now reserved for quantum computation. The fact that it was easy to reformulate in a GA way the Deutsch-Jozsa problem is very encourageing.

Finally, let us mention that certain attempts of using GA for the purposes of quantum computation can be found in the literature \cite{Somaroo,BDM3,HD}. Still, it seems that the approaches discussed so far reduce GA to the level of unitary operations or density matrices, i.e. objects that have a natural operator representation. In our approach, even ``pure states" are represented by elements of GA. This is why we can perform operations on pure states that have no couterpart in standard quantum computation. In this sense our geometric algorithm may be regarded as something conceptually in-between quantum algorithms and HRRs or BSC.

\section{Original Deutsch-Jozsa algorithm}

We assume there exists an oracle performing
\be
U_f|x\rangle|y\rangle
&=&
|x\rangle|y\oplus f(x)\rangle
\ee
where $f(x)\in\{0,1\}$. Now
\be
U_f|x\rangle\Big(|0\rangle-|1\rangle\Big)
&=&
|x\rangle\Big(|0\oplus f(x)\rangle-|1\oplus f(x)\rangle\Big)\\
&=&
|x\rangle\Big(|f(x)\rangle-|\neg f(x)\rangle\Big)
\ee
If $f(x)=0$ then
\be
U_f|x\rangle\Big(|0\rangle-|1\rangle\Big)
&=&
|x\rangle\Big(|0\rangle-|1\rangle\Big)
\ee
If $f(x)=1$ then
\be
U_f|x\rangle\Big(|0\rangle-|1\rangle\Big)
&=&
|x\rangle\Big(|1\rangle-|0\rangle\Big)
=
-|x\rangle\Big(|0\rangle-|1\rangle\Big)
\ee
The two cases imply 
\be
U_f|x\rangle\Big(|0\rangle-|1\rangle\Big)
&=&
(-1)^{f(x)}|x\rangle\Big(|0\rangle-|1\rangle\Big)
\ee
The Hadamard gate acts as follows
\be
U_H|0\rangle
&=&
\frac{1}{\sqrt{2}}
\Big(|0\rangle+|1\rangle\Big)\\
U_H|1\rangle
&=&
\frac{1}{\sqrt{2}}
\Big(|0\rangle-|1\rangle\Big)
\ee
Let 
\be
U_{n+1}=\underbrace{U_H\otimes\dots\otimes U_H}_{n+1}
\ee
Then
\be
U_{n+1}|\underbrace{0\dots 0}_{n}1\rangle
&=&
\frac{1}{\sqrt{2^{n+1}}}
\sum_{A_1\dots A_n=0}^1 |A_1\dots A_n\rangle\Big(|0\rangle-|1\rangle\Big)\\
&=&
\frac{1}{\sqrt{2^{n+1}}}
\sum_{A_1\dots A_{n+1}=0}^1 (-1)^{A_{n+1}}|A_1\dots A_n,A_{n+1}\rangle\label{1}\\
&=&
\frac{1}{\sqrt{2^{n+1}}}
\sum_{x=0}^{2^n-1} |x\rangle\Big(|0\rangle-|1\rangle\Big)\\
U_fU_{n+1}|\underbrace{0\dots 0}_{n}1\rangle
&=&
\frac{1}{\sqrt{2^{n+1}}}
\sum_{x=0}^{2^n-1} (-1)^{f(x)}|x\rangle\Big(|0\rangle-|1\rangle\Big)\\
U_{n+1}U_fU_{n+1}|\underbrace{0\dots 0}_{n}1\rangle
&=&
\frac{1}{\sqrt{2^{n}}}
\sum_{x=0}^{2^n-1} (-1)^{f(x)}U_n|x\rangle|1\rangle\\
&=&
\frac{1}{\sqrt{2^{n}}}
\sum_{A_1\dots A_n=0}^1 (-1)^{f(A_1\dots A_n)}U_n|A_1\dots A_n\rangle|1\rangle\\
&=&
\frac{1}{2^{n}}
\sum_{A_1\dots A_n=0}^1\sum_{B_1\dots B_n=0}^1 (-1)^{f(A_1\dots A_n)}
(-1)^{\sum_{k=1}^n A_kB_k}
|B_1\dots B_n\rangle|1\rangle\\
&=&
\frac{1}{2^{n}}
\sum_{B_1\dots B_n=0}^1\Big(\sum_{A_1\dots A_n=0}^1 (-1)^{f(A_1\dots A_n)}
(-1)^{\sum_{k=1}^n A_kB_k}\Big)
|B_1\dots B_n\rangle|1\rangle\\
&=&
\frac{1}{2^{n}}
\sum_{(B_1\dots B_n)\neq (0_1\dots 0_n)}\Big(\sum_{A_1\dots A_n=0}^1 (-1)^{f(A_1\dots A_n)}
(-1)^{\sum_{k=1}^n A_kB_k}\Big)
|B_1\dots B_n\rangle|1\rangle
\nonumber\\
&\pp= &
+
\frac{1}{2^{n}}
\sum_{A_1\dots A_n=0}^1 (-1)^{f(A_1\dots A_n)}
|0_1\dots 0_n\rangle|1\rangle
\\
&=&
\dots+
\frac{1}{2^{n}}
\sum_{x=0}^{2^n-1} (-1)^{f(x)}
|0_1\dots 0_n\rangle|1\rangle
\ee
If $f(x)=f(0)$ for any $x$ then
\be
{\rm RHS}
&=&
\dots+
\frac{1}{2^{n}}(-1)^{f(0)}
\sum_{x=0}^{2^n-1} 
|0_1\dots 0_n\rangle|1\rangle
=
(-1)^{f(0)}
|0_1\dots 0_n\rangle|1\rangle
\ee
If $f$ is balanced then
\be
{\rm RHS}
&=&
\dots+
\frac{1}{2^{n}}
\sum_{x=0}^{2^n-1} 
(-1)^{f(x)}|0_1\dots 0_n\rangle|1\rangle
\\
&=&
\frac{1}{2^{n}}
\sum_{(B_1\dots B_n)\neq (0_1\dots 0_n)}\Big(\sum_{A_1\dots A_n=0}^1 (-1)^{f(A_1\dots A_n)}
(-1)^{\sum_{k=1}^n A_kB_k}\Big)
|B_1\dots B_n\rangle|1\rangle
\ee
It is sufficient to look at the $|0\dots 01\rangle$ component to see if $f$ is constant or balanced.

\section{Geometric Algebra and its binary parametrization}

Here and in the next section we repeat the presentation from \cite{ACDM}. 

Euclidean-space GA is constructed as follows. One takes an $n$-dimensional linear space with orthonormal basis $\{e_1,\dots,e_n\}$. Directed subspaces are then associated with the set
\be
\{1,e_1,\dots,e_n,e_{12},e_{13}\dots,e_{n-1,n},\dots,e_{12\dots n}\}.
\ee
Here 1 corresponds to scalars, i.e. a 0-dimensional space. Then we have vectors (oriented segments), bivectors (oriented parallelograms), and so on. There exists a natural parametrization: $1=e_{0\dots 0}$, 
$e_1=e_{10\dots 0}$, $e_2=e_{010\dots 0}$, $\dots$, 
$e_{125}=e_{110010\dots 0}$, $\dots$, $e_{12\dots n-1,n}=e_{11\dots 1}$, which shows that there is a one-to-one relation between an $n$-bit number and an element of GA. An element with $k$ 1s and $n-k$ 0s is called a $k$-blade. 

A {\it geometric product\/} of $k$ 1-blades is a $k$-blade. For example, 
$e_{1248}=e_{1}e_{2}e_{4}e_{8}$. Moreover, $e_ne_m=-e_me_n$, if $m\neq n$, and $e_ne_n=1$, for any $n$. GA is a Clifford algebra \cite{BT} enriched by certain geometric interpretations and operations.

Particularly interesting is the form of the geometric product that occurs in the binary parametrization. Let us work out a few examples:
\be
e_1e_1
&=&
e_{10\dots 0}e_{10\dots 0}
=
1=e_{0\dots 0}=e_{(10\dots 0)\oplus (10\dots 0)}\\
e_1e_{12}
&=&
e_{10\dots 0}e_{110\dots 0}
=
e_1e_1e_2=e_2=e_{010\dots 0}=e_{(10\dots 0)\oplus (110\dots 0)}\\
e_{12}e_1
&=&
e_{110\dots 0}e_{10\dots 0}
=
e_1e_2e_1=-e_2e_1e_1=-e_2=-e_{010\dots 0}=-e_{(110\dots 0)\oplus(10\dots 0)}\\
e_{1257}e_{26}
&=&
e_{11001010\dots 0}e_{0100010\dots 0}
=
e_1e_2e_5e_7e_2e_6
=
(-1)^2e_1e_2e_2e_5e_7e_6
=
(-1)^2(-1)^1e_1e_2e_2e_5e_6e_7\nonumber\\
&=&
(-1)^3e_1e_5e_6e_7
=
(-1)^3
e_{10001110\dots 0}
=
(-1)^D
e_{(11001010\dots 0)\oplus(0100010\dots 0)}.
\ee
The number $D$ is the number of times a 1 from the right string had to ``jump" over a 1 from the left one during the process of shifting the right string to the left.
Symbolically the operation can be represented as
\be
\left[
\begin{array}{rl}
\longleftarrow & 01000100\dots 0\\
11001010\dots 0 & 
\end{array}
\right]
\mapsto
(-1)^D
\left[
\begin{array}{l}
01000100\dots 0\\
11001010\dots 0 
\end{array}
\right]
\mapsto
(-1)^D
\left[
\begin{array}{c}
01000100\dots 0\\
\oplus\\
11001010\dots 0 
\end{array}
\right]
=
(-1)^D
\left[
\begin{array}{l}
10001110\dots 0
\end{array}
\right]\nonumber
\ee
The above observations, generalized to arbitrary strings of bits, yield
\be
e_{A_1\dots A_n}e_{B_1\dots B_n}
&=&
(-1)^{\sum_{k<l}B_kA_l}e_{(A_1\dots A_n)\oplus(B_1\dots B_n)}.
\label{GAr}
\ee
Indeed, for two arbitrary strings of bits we have
\be
\left[
\begin{array}{rl}
\longleftarrow & B_1B_2\dots B_n\\
A_1A_2\dots A_n & 
\end{array}
\right]
\mapsto
(-1)^D
\left[
\begin{array}{l}
B_1B_2\dots B_n\\
A_1A_2\dots A_n 
\end{array}
\right]
\ee
where 
\be
D=B_1(A_2+\dots+A_n)+B_2(A_3+\dots+A_n)+\dots+B_{n-1}A_n=\sum_{k<l}B_kA_l.
\ee
\section{Cartan representation}

In this section we give an explicit matrix representation of GA. We begin with Pauli's matrices 
\be
\sigma_1
=
\left(
\begin{array}{cc}
0 & 1\\
1 & 0
\end{array}
\right),
\quad
\sigma_2
=
\left(
\begin{array}{cc}
0 & -i\\
i & 0
\end{array}
\right)
,
\quad
\sigma_3
=
\left(
\begin{array}{cc}
1 & 0\\
0 & -1
\end{array}
\right).
\ee
GA of a plane is represented as follows:  $1=2\times 2$ unit matrix, $e_1=\sigma_1$, 
$e_2=\sigma_2$, $e_{12}=\sigma_1\sigma_2=i\sigma_3$. Alternatively, we can write
$e_{00}=1$, $e_{10}=\sigma_1$, $e_{01}=\sigma_2$, 
$e_{11}=i\sigma_3$, and
\be
\alpha_{00} e_{00}+\alpha_{10} e_{10}+\alpha_{01} e_{01}+\alpha_{11} e_{11}
=
\left(
\begin{array}{cc}
\alpha_{00} +i\alpha_{11} & \alpha_{10} -i\alpha_{01}\\
\alpha_{10} +i\alpha_{01} & \alpha_{00}-i\alpha_{11}
\end{array}
\right).
\ee
This is equivalent to encoding $2^2=4$ real numbers into two complex numbers. 

In 3-dimensional space we have
$1=2\times 2$ unit matrix, $e_1=\sigma_1$, 
$e_2=\sigma_2$, $e_{3}=\sigma_3$, $e_{12}=\sigma_1\sigma_2=i\sigma_3$, 
$e_{13}=\sigma_1\sigma_3=-i\sigma_2$, 
$e_{23}=\sigma_2\sigma_3=i\sigma_1$, 
$e_{123}=\sigma_1\sigma_2\sigma_3=i$.

Now the representation of 
\be
\sum_{ABC=0,1}\alpha_{ABC}e_{ABC}
=
\left(
\begin{array}{cc}
\alpha_{000} +i\alpha_{111} + \alpha_{001}+i \alpha_{110},
& 
\alpha_{100}+i\alpha_{011} -i\alpha_{010}-\alpha_{101}\\
\alpha_{100}+i\alpha_{011} +i\alpha_{010}+\alpha_{101},
& 
\alpha_{000}+i\alpha_{111}
-\alpha_{001}-i \alpha_{110}
\end{array}
\right)
\ee
is equivalent to encoding $2^3=8$ real numbers into 4 complex numbers.

An arbitrary $n$-bit record can be encoded into the matrix algebra known as Cartan's representation of Clifford argebras \cite{BT}:
\be
e_{2k}
&=&
\underbrace{\sigma_1\otimes\dots\otimes \sigma_1}_{n-k}
\otimes\,\sigma_2\otimes
\underbrace{1\otimes\dots\otimes 1}_{k-1},\\
e_{2k-1}
&=&
\underbrace{\sigma_1\otimes\dots\otimes \sigma_1}_{n-k}
\otimes\,\sigma_3\otimes
\underbrace{1\otimes\dots\otimes 1}_{k-1}.
\ee

\section{GA formulation of the Deutsch-Jozsa problem}

Consider an $(n+1)$-dimensional Euclidean space with orthonormal basis $\{e_1,\dots e_{n+1}\}$, and its associated GA.
The basis vector $e_{n+1}$ in binary parametrization corresponds to $e_{0\dots 01}$. Recall that 
\be
e_{A_1\dots A_{n+1}}e_{B_1\dots B_{n+1}}=(-1)^{\sum_{i<j}B_iA_j}e_{(A_1\dots A_{n+1})\oplus(B_1\dots B_{n+1})}
\ee
and, in particular,
\be
e_{A_1\dots A_{n+1}}e_{0\dots 0 1}
&=&
e_{A_1\dots A_n,A_{n+1}\oplus 1}\\
e_{A_1\dots A_{n+1}}e_{0\dots 0 10}
&=&
(-1)^{A_{n+1}}
e_{A_1\dots A_{n-1},A_n\oplus 1,A_{n+1}}
\ee
Consider
\be
E_{n+1}
&=&
\sum_{A_1\dots A_{n+1}=0}^1e_{A_1\dots A_{n+1}}\\
E_{n+1} e_{0\dots 0 10}
&=&
\sum_{A_1\dots A_{n+1}=0}^1e_{A_1\dots A_{n+1}}e_{0\dots 0 10}\\
&=&
\sum_{A_1\dots A_{n+1}=0}^1(-1)^{A_{n+1}}
e_{A_1\dots A_{n-1},A_n\oplus 1,A_{n+1}}\\
&=&
\sum_{A_1\dots A_{n-1}A_{n+1}=0}^1(-1)^{A_{n+1}}
e_{A_1\dots A_{n-1},0\oplus 1,A_{n+1}}
+
\sum_{A_1\dots A_{n-1}A_{n+1}=0}^1(-1)^{A_{n+1}}
e_{A_1\dots A_{n-1},1\oplus 1,A_{n+1}}\\
&=&
\sum_{A_1\dots A_{n-1}A_{n+1}=0}^1(-1)^{A_{n+1}}
e_{A_1\dots A_{n-1},1,A_{n+1}}
+
\sum_{A_1\dots A_{n-1}A_{n+1}=0}^1(-1)^{A_{n+1}}
e_{A_1\dots A_{n-1},0,A_{n+1}}\\
&=&
\sum_{A_1\dots A_{n+1}=0}^1(-1)^{A_{n+1}}
e_{A_1\dots A_{n+1}}
\ee
The influence of $E_{n+1}$ on $e_{0\dots 0 10}$ is similar to (\ref{1}):
\be
E_{n+1} e_{0\dots 0 10}
&=&
\sum_{A_1\dots A_{n+1}=0}^1(-1)^{A_{n+1}}e_{A_1\dots A_{n+1}}\\
&=&
\sum_{A_1\dots A_{n}=0}^1
\Big(
e_{A_1\dots A_{n}0}
-
e_{A_1\dots A_{n}1}
\Big)
\\
U_{n+1}|\underbrace{0\dots 0}_{n}1\rangle
&=&
\frac{1}{\sqrt{2^{n+1}}}
\sum_{A_1\dots A_{n+1}=0}^1 (-1)^{A_{n+1}}|A_1\dots A_{n+1}\rangle\\
&=&
\frac{1}{\sqrt{2^{n+1}}}
\sum_{A_1\dots A_{n}=0}^1 
\Big(
|A_1\dots A_{n}0\rangle
-
|A_1\dots A_{n}1\rangle
\Big)
\ee
Now assume there exists an oracle $E_f$ that performs
\be
E_fe_{A_1\dots A_{n}A_{n+1}}
&=&
e_{A_1\dots A_{n},A_{n+1}\oplus f(A_1\dots A_{n})}
=
e_{A_1\dots A_{n},A_{n+1}}e_{0\dots 0,f(A_1\dots A_{n})}
\ee
Then
\be
E_fE_{n+1} e_{0\dots 0 10}
&=&
\sum_{A_1\dots A_{n}=0}^1
E_f\Big(
e_{A_1\dots A_{n}0}
-
e_{A_1\dots A_{n}1}
\Big)\\
&=&
\sum_{A_1\dots A_{n}=0}^1
E_f\Big(
e_{A_1\dots A_{n}0}
-
e_{A_1\dots A_{n}1}
\Big)\\
&=&
\sum_{A_1\dots A_{n}=0}^1
\Big(
e_{A_1\dots A_{n},f(A_1\dots A_{n})}
-
e_{A_1\dots A_{n},\neg f(A_1\dots A_{n})}
\Big)\\
&=&
\sum_{A_1\dots A_{n}=0}^1
(-1)^{f(A_1\dots A_{n})}
\Big(
e_{A_1\dots A_{n}0}
-
e_{A_1\dots A_{n}1}
\Big)
\ee
In GA there exists an operation of {\it reverse\/} which reverses the order as follows: If $X=e_1e_2\dots e_{k-1}e_k$ then the reverse of $X$ is 
\be
X^{\dag}=e_ke_{k-1}\dots e_2e_1 =(-1)^{k(k-1)/2}X
\ee
By linearity we extend it to all multivectors. In binary parametrization the number $k$ describes the number of 1s in 
$e_{A_1\dots A_{n+1}}$, i.e. $k=\sum_{j=1}^{n+1}A_j$. 
So consider
\be
F_{n+1}
&=&
\sum_{A_1\dots A_{n}=0}^1e_{A_1\dots A_{n}0}^{\dag}\\
&=&
\sum_{A_1\dots A_{n}=0}^1(-1)^{k(k-1)/2}e_{A_1\dots A_{n}0}
\ee
Here $k=\sum_{j=1}^{n}A_j$ since the last bit is 0. Now
\be
F_{n+1}E_fE_{n+1} e_{0\dots 0 10}
&=&
\sum_{A_1\dots A_{n}=0}^1
(-1)^{f(A_1\dots A_{n})}
F_{n+1}\Big(
e_{A_1\dots A_{n}0}
-
e_{A_1\dots A_{n}1}
\Big)\\
&=&
\sum_{A_1\dots A_{n}=0}^1
(-1)^{f(A_1\dots A_{n})}
F_{n+1}
e_{A_1\dots A_{n}0}
+\dots\\
&=&
\sum_{A_1\dots A_{n}=0}^1
(-1)^{f(A_1\dots A_{n})}
\sum_{B_1\dots B_{n}=0}^1(-1)^{k(k-1)/2}e_{B_1\dots B_{n}0}
e_{A_1\dots A_{n}0}
+\dots\\
&=&
\sum_{A_1\dots A_{n}=0}^1
(-1)^{f(A_1\dots A_{n})}
\sum_{B_1\dots B_{n}=0}^1(-1)^{k(k-1)/2}
(-1)^{\sum_{k<l}A_kB_l}
e_{(B_1\dots B_{n}0)\oplus(A_1\dots A_{n}0)}
+\dots\\
&=&
\sum_{A_1\dots A_{n}=0}^1
(-1)^{f(A_1\dots A_{n})}
e_{0\dots 0}
+\dots
\ee
The dots denote all those term where the binary indices contain at least one 1. The two powers of $-1$ have cancelled out since $e_{A_1\dots A_{n+1}}^{\dag}e_{A_1\dots A_{n+1}}=1=e_{0\dots 0}$. Finally
\be
F_{n+1}E_fE_{n+1} e_{0\dots 0 10}
&=&
\sum_{A_1\dots A_{n}=0}^1
(-1)^{f(A_1\dots A_{n})}
e_{0\dots 0}
+\dots
\ee
Now let $\Pi$ project on  $1=e_{0\dots 0}$. It follows that 
\be
\Tr{\Pi}(F_{n+1}E_fE_{n+1}e_{0\dots 0 10})
&=&
\sum_{A_1\dots A_{n}=0}^1
(-1)^{f(A_1\dots A_{n})}
\Tr e_{0\dots 0}
=
N\sum_{A_1\dots A_{n}=0}^1
(-1)^{f(A_1\dots A_{n})}\\
&=&
\left\{
\begin{array}{rl}
(-1)^{f(0\dots 0)} N2^{n} & {\rm if}\,f{\rm\, is\,constant}\\
0 & {\rm if}\,f{\rm\, is\,balanced}
\end{array}
\right.
\ee
Here $N=\Tr 1$ is the dimension of the representation of GA.
We have achieved the same goal as the quantum algorithm.

We have to point out at this moment a possible error one can make. 
Let us note that in the step
\be
E_fe_{A_1\dots A_{n}A_{n+1}}
&=&
e_{A_1\dots A_{n},A_{n+1}}e_{0\dots 0,f(A_1\dots A_{n})}
\ee
we have $E_f$ on the left and $e_{0\dots 0,f(A_1\dots A_{n})}$ on the right. It might appear that it would be simpler and more natural to write $E_f$ on the right as well. However, this would be misleading since
\be
F_{n+1}E_fe_{A_1\dots A_{n}A_{n+1}}
&=&
F_{n+1}\Big(e_{A_1\dots A_{n}A_{n+1}}e_{0\dots 0,f(A_1\dots A_{n})}\Big)
\\
&\neq&
\Big(F_{n+1}e_{A_1\dots A_{n}A_{n+1}}\Big)e_{0\dots 0,f(A_1\dots A_{n})}
=
E_fF_{n+1}e_{A_1\dots A_{n}A_{n+1}}
\ee

\section{Explicit examples}

\subsection{Two bits}

GA of a plane consists of:  $1=2\times 2$ unit matrix, $e_1=\sigma_1$, 
$e_2=\sigma_2$, $e_{12}=\sigma_1\sigma_2=i\sigma_3$. Alternatively, we can write
$e_{00}=1$, $e_{10}=\sigma_1$, $e_{01}=\sigma_2$, 
$e_{11}=i\sigma_3$. 
\be
\alpha_{00} e_{00}+\alpha_{10} e_{10}+\alpha_{01} e_{01}+\alpha_{11} e_{11}
=
\left(
\begin{array}{cc}
\alpha_{00} +i\alpha_{11} & \alpha_{10} -i\alpha_{01}\\
\alpha_{10} +i\alpha_{01} & \alpha_{00}-i\alpha_{11}
\end{array}
\right).
\ee
\be
E_2
=
\left(
\begin{array}{cc}
1 +i & 1 -i \\
1 +i & 1-i 
\end{array}
\right).
\ee
\be
F_2
&=&
e_{00}^{\dag}+ e_{10}^{\dag}
=
e_{00}+ e_{10}
=1+\sigma_1
=
\left(
\begin{array}{cc}
1 & 1\\
1 & 1
\end{array}
\right).
\ee
\be
E_2e_{10}
&=&
\Big(e_{00}+e_{10}+e_{01}+e_{11}\Big)e_{10}
=
e_{10}+e_{00}-e_{11}-e_{01}\\
&=&
\Big(1+\sigma_1+\sigma_2+i\sigma_3\Big)\sigma_1
=
\sigma_1+1-i\sigma_3-\sigma_2
\ee
\be
E_fE_2e_{10}
&=&
E_f\Big(e_{10}+e_{00}-e_{11}-e_{01}\Big)\\
&=&
e_{1,0\oplus f(1)}+e_{0,0\oplus f(0)}-e_{1,1\oplus f(1)}-e_{0,1\oplus f(0)}\\
&=&
e_{1,f(1)}+e_{0,f(0)}-e_{1,\neg f(1)}-e_{0,\neg f(0)}
\ee

\subsubsection{Case $f(0)=f(1)=0$}

\be
E_fE_2e_{10}
&=&
e_{1,f(1)}+e_{0,f(0)}-e_{1,\neg f(1)}-e_{0,\neg f(0)}\\
&=&
e_{10}+e_{00}-e_{11}-e_{01}\\
&=&
\sigma_1+1-i\sigma_3-\sigma_2\\
&=&
\left(
\begin{array}{cc}
1-i & 1+i\\
1-i & 1+i
\end{array}
\right)
\ee
\be
F_2E_fE_2e_{10}
&=&
\left(
\begin{array}{cc}
1 & 1\\
1 & 1
\end{array}
\right)
\left(
\begin{array}{cc}
1-i & 1+i\\
1-i & 1+i
\end{array}
\right)
=
2
\left(
\begin{array}{cc}
1-i & 1+i\\
1-i & 1+i
\end{array}
\right)
\ee
\be
\Tr{\Pi} F_2E_fE_2e_{10}
=4
\ee

\subsubsection{Case $f(0)=f(1)=1$}

\be
E_fE_2e_{10}
&=&
e_{1,f(1)}+e_{0,f(0)}-e_{1,\neg f(1)}-e_{0,\neg f(0)}\\
&=&
e_{11}+e_{01}-e_{10}-e_{00}
\ee
Since this is minus the result from the previus subsection, we immediately get
\be
\Tr{\Pi} F_2E_fE_2e_{10}
=-4
\ee

\subsubsection{Case $f(0)=0$, $f(1)=1$}

\be
E_fE_2e_{10}
&=&
e_{1,f(1)}+e_{0,f(0)}-e_{1,\neg f(1)}-e_{0,\neg f(0)}\\
&=&
e_{11}+e_{00}-e_{10}-e_{01}\\
&=&
i\sigma_3+1-\sigma_1-\sigma_2\\
&=&
\left(
\begin{array}{cc}
1+i & -1+i\\
-1-i & 1-i
\end{array}
\right)
\ee
\be
F_2E_fE_2e_{10}
&=&
\left(
\begin{array}{cc}
1 & 1\\
1 & 1
\end{array}
\right)
\left(
\begin{array}{cc}
1+i & -1+i\\
-1-i & 1-i
\end{array}
\right)
=
\left(
\begin{array}{cc}
0 & 0\\
0 & 0
\end{array}
\right)
\ee
\be
\Tr{\Pi} F_2E_fE_2e_{10}
=0
\ee
Alternatively
\be
F_2E_fE_2e_{10}
&=&
(1+\sigma_1)
(i\sigma_3+1-\sigma_1-\sigma_2)
\\
&=&
(i\sigma_3+1-\sigma_1-\sigma_2)
+
(i\sigma_1\sigma_3+\sigma_1-1-\sigma_1\sigma_2)\\
&=&
(i\sigma_3+1-\sigma_1-\sigma_2)
+
(i(-i\sigma_2)+\sigma_1-1-i\sigma_3)=0
\ee
\subsubsection{Case $f(0)=1$, $f(1)=0$}

\be
E_fE_2e_{10}
&=&
e_{1,f(1)}+e_{0,f(0)}-e_{1,\neg f(1)}-e_{0,\neg f(0)}\\
&=&
e_{10}+e_{01}-e_{11}-e_{00}
\ee
This is minus the result from the previou section and therefore $\Tr{\Pi} F_2E_fE_2e_{10}=0$. 

Summing up, constant functions were producing $2(-1)^{f(0)}2^1$, and balanced functions implied 0, as it should be on general grounds.

\subsection{Three bits}

In 3-dimensional space we have
$1=2\times 2$ unit matrix, $e_1=\sigma_1$, 
$e_2=\sigma_2$, $e_{3}=\sigma_3$, $e_{12}=\sigma_1\sigma_2=i\sigma_3$, 
$e_{13}=\sigma_1\sigma_3=-i\sigma_2$, 
$e_{23}=\sigma_2\sigma_3=i\sigma_1$, 
$e_{123}=\sigma_1\sigma_2\sigma_3=i$. The operation $\Tr\Pi$ corresponds in this representation to taking the real part of trace (only $e_{000}=1$ and $e_{111}=i$ have nonzero trace).

Now the representation of a general element reads
\be
\sum_{ABC=0,1}\alpha_{ABC}e_{ABC}
=
\left(
\begin{array}{cc}
\alpha_{000} +i\alpha_{111} + \alpha_{001}+i \alpha_{110},
& 
\alpha_{100}+i\alpha_{011} -i\alpha_{010}-\alpha_{101}\\
\alpha_{100}+i\alpha_{011} +i\alpha_{010}+\alpha_{101},
& 
\alpha_{000}+i\alpha_{111}
-\alpha_{001}-i \alpha_{110}
\end{array}
\right)
\ee
\be
E_3
&=&
\sum_{ABC=0,1}e_{ABC}
=
\left(
\begin{array}{cc}
1 +i + 1+i ,
& 
1+i -i-1\\
1+i +i+1,
& 
1+i
-1-i 
\end{array}
\right)
=
2\left(
\begin{array}{cc}
1 +i,
& 
0\\
1+i ,
& 
0
\end{array}
\right)
\ee
\be
F_3
&=&
\sum_{AB=0,1}e_{AB0}^{\dag}
=
e_{000}^{\dag}
+
e_{100}^{\dag}
+
e_{010}^{\dag}
+
e_{110}^{\dag}
=
1+\sigma_1+\sigma_2+(\sigma_1\sigma_2)^{\dag}\\
&=&
1+\sigma_1+\sigma_2+\sigma_2\sigma_1
=
1+\sigma_1+\sigma_2-i\sigma_3
=
\left(
\begin{array}{cc}
1-i & 1-i\\
1+i & 1+i
\end{array}
\right)
\ee
\be
E_3e_{010}
&=&
2\left(
\begin{array}{cc}
1 +i
& 
0\\
1+i 
& 
0
\end{array}
\right)
\left(
\begin{array}{cc}
0&-i \\
i
& 
0
\end{array}
\right)
=
2
\left(
\begin{array}{cc}
0 &1-i \\
0 & 1-i
\end{array}
\right)\\
&=&
\Big(
e_{000}
+
e_{100}
+
e_{010}
+
e_{001}
+
e_{110}
+
e_{011}
+
e_{101}
+
e_{111}
\Big)e_{010}\\
&=&
e_{000}e_{010}
+
e_{100}e_{010}
+
e_{010}e_{010}
+
e_{001}e_{010}
+
e_{110}e_{010}
+
e_{011}e_{010}
+
e_{101}e_{010}
+
e_{111}e_{010}\\
&=&
e_{010}
+
e_{110}
+
e_{000}
-
e_{011}
+
e_{100}
-
e_{001}
-
e_{111}
-
e_{101}\\
&=&
\sigma_2
+
\sigma_1\sigma_2
+
1
-
\sigma_2\sigma_3
+
\sigma_1
-
\sigma_3
-
\sigma_1\sigma_2\sigma_3
-
\sigma_1\sigma_3\\
&=&
\sigma_2
+
i\sigma_3
+
1
-
i\sigma_1
+
\sigma_1
-
\sigma_3
-
i
+
i\sigma_2\\
&=&
\left(
\begin{array}{cc}
i +1-1-i,
& 
-i-i+1+1
\\
i-i+1-1,
& 
-i+1+1-i
\end{array}
\right)
\ee
\be
E_fE_3e_{010}
&=&
e_{01,0\oplus f(01)}
+
e_{11,0\oplus f(11)}
+
e_{00,0\oplus f(00)}
-
e_{01,1\oplus f(01)}
+
e_{10,0\oplus f(10)}
-
e_{00,1\oplus f(00)}
-
e_{11,1\oplus f(11)}
-
e_{10,1\oplus f(10)}\nonumber\\
&=&
e_{01,f(01)}
+
e_{11,f(11)}
+
e_{00,f(00)}
-
e_{01,\neg f(01)}
+
e_{10,f(10)}
-
e_{00,\neg f(00)}
-
e_{11,\neg f(11)}
-
e_{10,\neg f(10)}\nonumber\\
&=&
e_{00,f(00)}
-
e_{00,\neg f(00)}
+
e_{01,f(01)}
-
e_{01,\neg f(01)}
+
e_{10,f(10)}
-
e_{10,\neg f(10)}
+
e_{11,f(11)}
-
e_{11,\neg f(11)}
\ee
\subsubsection{Case of constant $f$, $f(00)=0$}
\be
E_fE_3e_{010}
&=&
e_{000}
-
e_{001}
+
e_{010}
-
e_{011}
+
e_{100}
-
e_{101}
+
e_{110}
-
e_{111}\\
&=&
1
-
\sigma_3
+
\sigma_2
-
i\sigma_1
+
\sigma_1
+
i\sigma_2
+
i\sigma_3
-
i
\\
&=&
\left(
\begin{array}{cc}
1-1+i-i,
& 
-i-i+1+1\\
i-i+1-1,
& 
1+1-i-i
\end{array}
\right)
=
2(1-i)
\left(
\begin{array}{cc}
0
& 
1\\
0
& 
1\end{array}
\right)
\ee
\be
F_3E_fE_3e_{010}
&=&
\left(
\begin{array}{cc}
1-i & 1-i\\
1+i & 1+i
\end{array}
\right)
2(1-i)
\left(
\begin{array}{cc}
0
& 
1\\
0
& 
1\end{array}
\right)\\
&=&
2(1-i)
\left(
\begin{array}{cc}
1-i & 1-i\\
1+i & 1+i
\end{array}
\right)
\left(
\begin{array}{cc}
0
& 
1\\
0
& 
1\end{array}
\right)\\
&=&
2(1-i)
\left(
\begin{array}{cc}
0
& 
2(1-i)\\
0
& 
2(1+i)\end{array}
\right)
\ee
\be
\Re{\Tr} F_3E_fE_3e_{010}
&=&
2(1-i)2(1+i)
=8=2(-1)^02^2
\ee
\subsubsection{Case of constant $f$, $f(00)=1$}
\be
E_fE_3e_{010}
&=&
e_{001}
-
e_{000}
+
e_{011}
-
e_{010}
+
e_{101}
-
e_{100}
+
e_{111}
-
e_{110}
\ee
\be
\Re{\Tr} F_3E_fE_3e_{010}
&=&
-2(1-i)2(1+i)
=-8=2(-1)^12^2
\ee
\subsubsection{Case of balanced $f$, $f(00)=0$, $f(10)=0$}

\be
E_fE_3e_{010}
&=&
e_{00,f(00)}
+
e_{10,f(10)}
+
e_{01,f(01)}
+
e_{11,f(11)}
-
e_{00,\neg f(00)}
-
e_{10,\neg f(10)}
-
e_{01,\neg f(01)}
-
e_{11,\neg f(11)}
\nonumber\\
&=&
e_{000}
+
e_{100}
+
e_{011}
+
e_{111}
-
e_{001}
-
e_{101}
-
e_{010}
-
e_{110}
\nonumber\\
&=&
e_{000}
+
e_{100}
-
e_{010}
-
e_{001}
+
e_{011}
-
e_{101}
-
e_{110}
+
e_{111}
\nonumber\\
&=&
\left(
\begin{array}{cc}
1 +i -1 - i ,
& 
1+i +i+1\\
1+i -i-1,
& 
1+i+1+i
\end{array}
\right)
=
2(1+i)
\left(
\begin{array}{cc}
0 ,
& 
1\\
0,
& 
1
\end{array}
\right)
\ee
\be
F_3E_fE_3e_{010}
&=&
2(1+i)
\left(
\begin{array}{cc}
0
& 
2(1-i)\\
0
& 
2(1+i)\end{array}
\right)
\ee
\be
\Re\Tr
F_3E_fE_3e_{010}
&=&
\Re 2(1+i)2(1+i)=\Re 8i=0
\ee
\section{Final remarks}

The above examples show that GA allows for a host of new mathematical tricks with respect to standard quantum computation. The representations of binary numbers are different. There is no distinction between ``state vectors" and ``operators". One can multiply ``state vectors" without increasing the dimension. In the above examples both 2-bit and 3-bit problems were represented by $2\times 2$ matrices, a fact showing that one may expect GA to involve less redundancy than standard tensor representations. One can speak of entanglement in GA representations even though the ``states" are not tensored with one another. Here again one finds close analogies to what is known from HRRs and BSC. And, last but not least, it seems there is no general difficulty with translating quantum operations into GA forms, and one can expect all quantum algorithms to have GA analogues.

\end{document}